\documentclass[12pt]{elsarticle}

\usepackage{epsfig}

\input{symbol}

\newcommand{\BDDK}{\ensuremath{B\to \Dbar^{(*)} D^{(*)} K}}
\newcommand{\BDDsK}{\ensuremath{B\to \Dbar D^* K}}
\newcommand{\BDsDK}{\ensuremath{B\to \Dbar^* D K}}
\newcommand{\BDsDsK}{\ensuremath{B\to \Dbar^{*} D^{*} K}}
\newcommand{\BDDKspec}{B \rightarrow \Dbar D  K  }
\newcommand{\bccs}{b \rightarrow c \overline{c} s  }

\def\modex{\ensuremath{\Bz\to \Dstarm \Dstarp K^0}}
\def\modexi{\ensuremath{\Bu\to \Dzb \Dz \Kp}}

\newcommand\CellTop{\rule{0pt}{2.35ex}}

\journal{Physics Letters B}

\begin{document}

\begin{frontmatter}

\title {A precise isospin analysis of $\BDDK$ decays}

\author{Vincent Poireau}
\address{Laboratoire d'Annecy-le-Vieux de Physique des Particules (LAPP), Universit\'e de Savoie, CNRS/IN2P3, F-74941 Annecy-Le-Vieux, France}
\author{Marco Zito}
\address{DSM/Irfu/SPP, CEA-Saclay, 91191 Gif/Yvette, France}

\begin{abstract}
We present a precise isospin analysis of the $\BDDK$ decays using new recent experimental measurements on these final states. The decays $\BDDK$,  originating from $\bccs$ transitions, are linked by a rich set of isospin properties. The isospin relations that connect the decay modes are presented and a fit is performed to obtain the isospin amplitudes and phases. We discuss the results of the fit and present a new measurement of the ratio of branching fractions $\BR(\Upsilon(4S) \rightarrow B^+ B^-)$ and $\BR(\Upsilon(4S) \rightarrow \Bz \Bzb)$. We finally discuss the implications of our findings for the measurement of the unitarity matrix parameters
$\sin(2 \beta)$ and $\cos(2 \beta)$ using these decays.
\end{abstract}

\end{frontmatter}

\section{Introduction}

In this Letter, we use an isospin analysis to establish relations between the different $\BDDK$ decays. These decays proceed via $\bccs$ transitions, which are known to present peculiar isospin properties~\cite{ref:sanda}.  The possibility that a large fraction of $\bccs$  decays hadronize as $\BDDK$ was first suggested in Ref.~\cite{ref:buchalla} in the context of the discrepancy between the
measured $B$ semi-leptonic rate and the theoretical prediction. This hypothesis was confirmed by many experimental results where it was found that $\BDDK$ decays account for about $4\%$ of the $B^0$ and $B^+$ decays~\cite{ref:cleoaleph,ref:ddk2003,ref:ddk2011}. These results provide the input for the isospin analysis and the test of the isospin relations. An additional motivation for an in-depth study of
these channels is the possibility, originally discussed in Refs.~\cite{ref:CPDDK_1,ref:CPDDK_2,ref:CPDDK_3}, to measure $\sin(2 \beta)$ and $\cos(2 \beta)$ using these decays. Indeed they proceed through the same quark current than the gold-plated mode $\Bz \to J/\Psi \Kz$ and are not Cabibbo-suppressed to the difference of the $\Bz \to \Dbar^{(*)} D^{(*)}$ modes.

This Letter, which updates and supersedes a previous investigation reported
in Ref.~\cite{ref:previous}, presents the complete set of isospin relations for
$\BDDK$ decays. They are compared to the measurements through a fit
of the experimental data which determines the isospin amplitudes.
There are 22 possible modes for the $\BDDK$ decays; here the $B$ is either a $B^0$ or a $B^+$, the $D^{(*)}$ is either a $D^0$, $D^{*0}$, $D^+$, or $D^{*+}$, the $\Dbar^{(*)}$ is the charge conjugate of $D^{(*)}$, and the $K$ is either a $K^+$ or a $K^0$.

These decays have been the object of many  experimental investigations during the past years. In particular, the BABAR Collaboration~\cite{ref:ddk2011} published recently a complete set of measurements of the 22 branching fractions with an excellent accuracy. They used $471 \times 10^6~B\Bbar$ events collected at the $\Upsilon(4S)$ resonance, corresponding to an integrated luminosity of $429~\mathrm{fb}^{-1}$.
The Belle Collaboration performed a measurement of the branching fractions of the modes \modex~\cite{ref:belleDDKCP} and \modexi~\cite{ref:belleDDKDsJ} using $449 \times 10^6~B\Bbar$ pairs. All these results are used in our analysis.

With respect to the previous study~\cite{ref:previous}, the statistical and systematic precision
on the experimental data is improved by a factor three or larger,
thereby improving by the same amount the statistical power of the tests
performed. This allows to put on a firm ground the conclusion that we draw from
this study.

In addition to the higher statistics, another improvement of the analysis shown in this Letter is the fact that the branching ratios $\BR(\Upsilon(4S) \rightarrow B^+ B^-)$ and
$\BR(\Upsilon(4S) \rightarrow \Bz \Bzb)$, needed to compare the neutral to charged $B$ meson
decays measured at an $e^+ e^-$ machine operating
at the $\Upsilon(4S)$ resonance, are presently known with a good accuracy. This good knowledge of the ratio helps to constrain more strongly the fit performed here.

The aim of this study is:
\begin{itemize}
\item to verify the isospin relations using a new set
of precise experimental results;
\item to provide some insight into the $\BDDK$ decay mechanism
from the inspection of the isospin
amplitudes;
\item to discuss the implications of our findings for the
measurement of
$\sin(2 \beta)$ and
$\cos(2 \beta)$ using these decays.
\end{itemize}

It has to be noted that other authors~\cite{ref:eriksson} studied $\BDDK$ decays in the context of color rearrangement models, and compared their predictions with the experimental measurements.

\section{Isospin relations for $\BDDK$ decays}

A full derivation and discussion of the isospin relations for these decays
can be found in Ref.~\cite{ref:previous}. Here only the main results
is summarized.

\begin{figure}[htb]
\begin{minipage}{8cm}
\begin{flushleft}
\includegraphics[width=6cm]{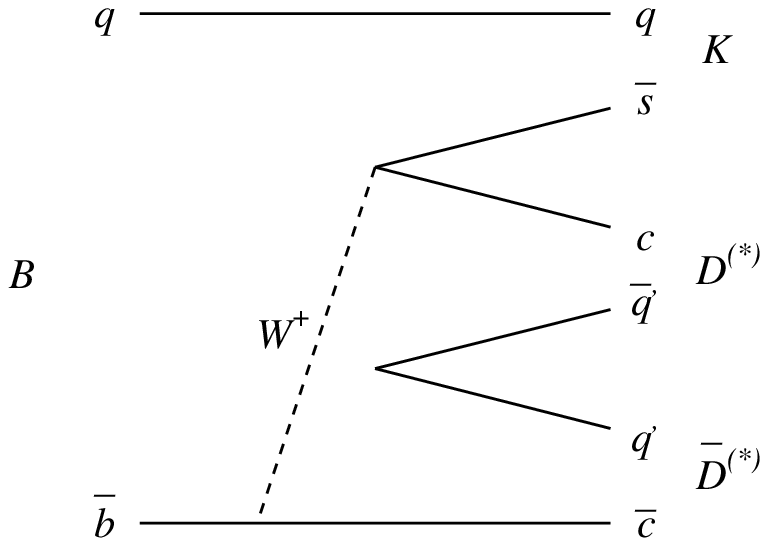}\\
\end{flushleft}
\label{fig:dstpi-diagrams-1}
\end{minipage}
\hfill
\begin{minipage}{8cm}
\begin{flushright}

\includegraphics[width=6cm]{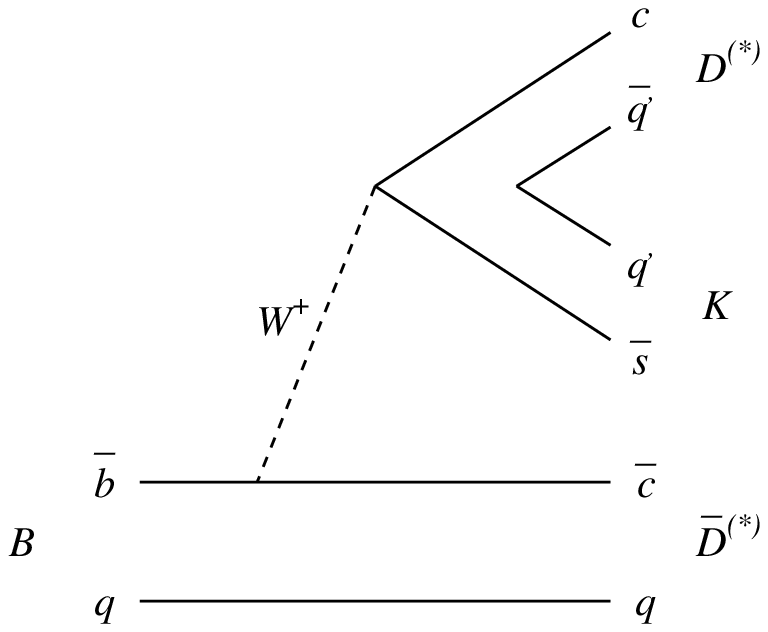}\\
\end{flushright}
\label{fig:dstpi-diagrams-2}
\end{minipage}
\caption{Left: internal \W-emission diagram for the decays
$B \to \Dbar^{(*)} D^{(*)} K$.
Right: external \W-emission diagram for the decays
$B \to \Dbar^{(*)} D^{(*)} K$.
  }
\label{Fi:diagrams}
\end{figure}

\begin{figure}[htb]
\begin{minipage}{8cm}
\begin{flushleft}
\includegraphics[width=6cm]{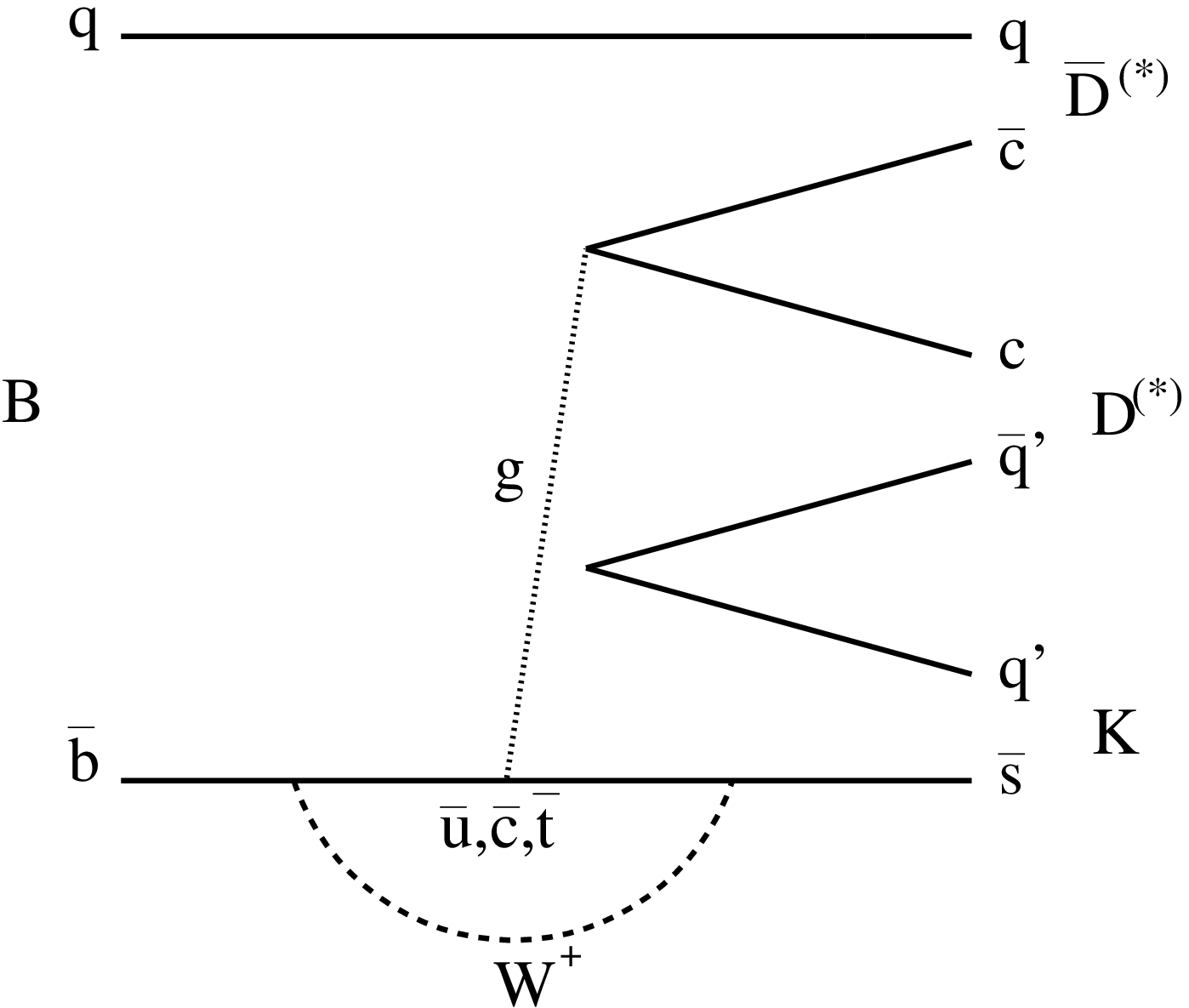}\\
\end{flushleft}
\end{minipage}
\hfill
\begin{minipage}{8cm}
\begin{flushright}
\includegraphics[width=6cm]{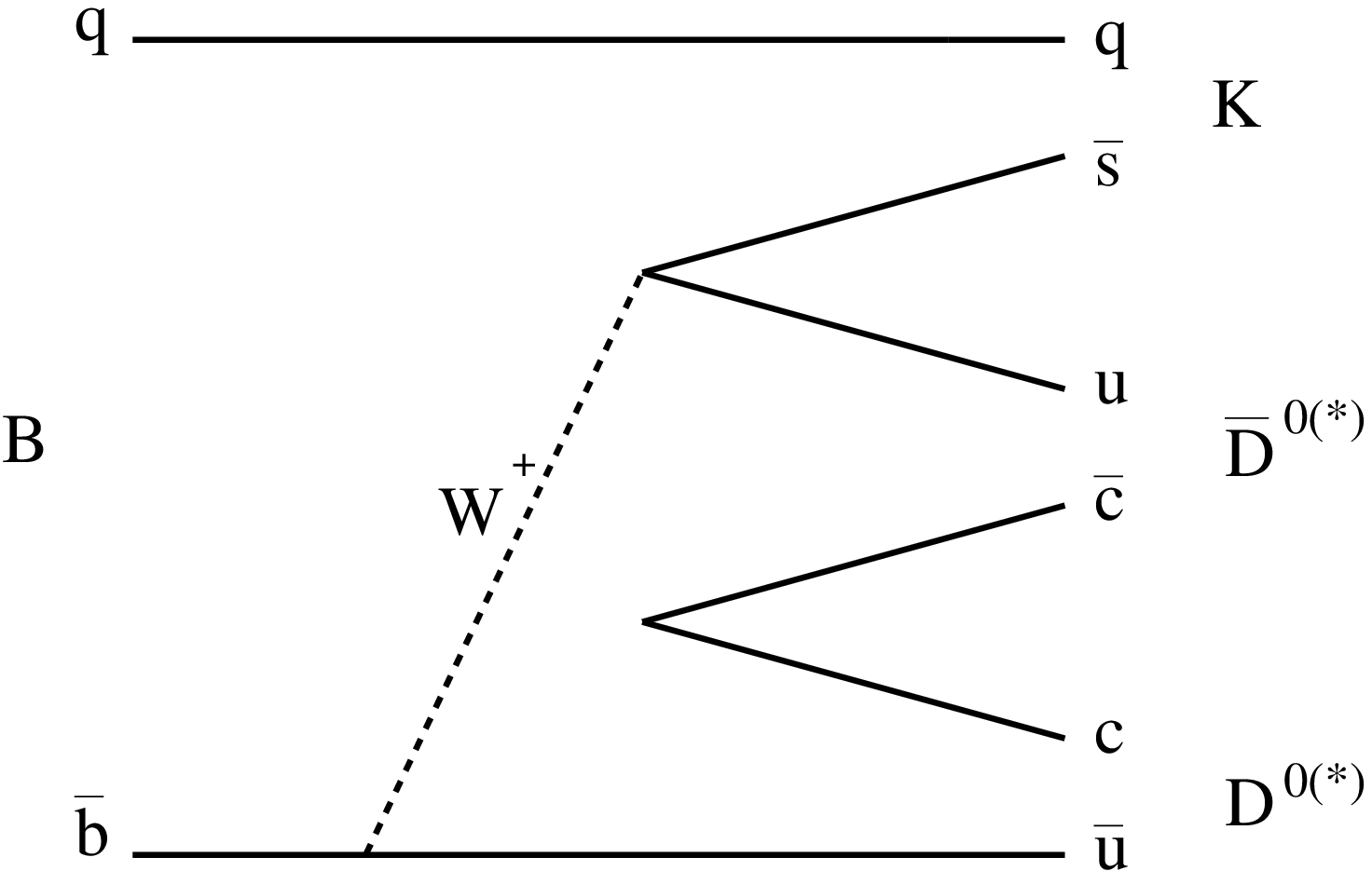}\\
\end{flushright}
\end{minipage}
\caption{Left: QCD penguin diagram  for the decays
$B \to \Dbar^{(*)} D^{(*)} K$.
Right: Cabibbo-suppressed diagram with $\Delta I = 1$ amplitude.}
\label{fig:supDiagram}
\end{figure}

The $\BDDK$ decays proceed via a  $\bccs$ current through the diagrams of
Fig.~\ref{Fi:diagrams}. Depending on the final state, the external W-emission
diagram, the internal W-emission diagram (which is color-suppressed), or both
contribute to the transition amplitude.
A penguin diagram, shown in Fig.~\ref{fig:supDiagram} (left plot),
can also contribute to the $\bccs$ current. It is expected to be
suppressed relatively to the tree diagrams of Fig.~\ref{Fi:diagrams}
and does not modify the isospin relations.

The decays $\Bz \to \Dbar^{(*)0} D^{(*)0} \Kz$
and $\Bp \to \Dbar^{(*)0} D^{(*)0} \Kp$
could also proceed through a different diagram,
shown in Fig.~\ref{fig:supDiagram} (right plot), which could introduce a
$\Delta I = 1$ amplitude. However this diagram proceeds through two
suppressed weak vertices $b \rightarrow u W$ and
 $W \rightarrow s \bar u$ and a $c \bar c$ pair must be
extracted from the vacuum, instead of a light quark pair as
in the Cabibbo-allowed diagrams. This amplitude is therefore
suppressed by at least a factor $\lambda^2$, where
$\lambda$ is the expansion parameter of the Wolfenstein parametrization.
For these reasons we expect that $\Delta I = 0$
holds to  an excellent precision.

As already mentioned, the isospin properties of the $\bccs$ current
are well known and follow from the fact that only isoscalar quarks
are involved. Therefore this is a $\Delta I = 0$ weak transition
and the final state is an isospin eigenstate.

The isospin properties translate in the following set of relations~\cite{ref:previous}
\begin{eqnarray}
A(\Bz \rightarrow D^- \Dz K^+ ) &=& \frac{1} {\sqrt{6}}  A_1 -\frac{1} {\sqrt{2}}
A_0 \label{eq:isorel01}\\
A(\Bz \rightarrow D^- D^+ \Kz ) &=& \frac{1} {\sqrt{6}} A_1 +\frac{1} {\sqrt{2}}
A_0 \label{eq:isorel02}\\
A(\Bz \rightarrow \Dzb  \Dz \Kz )& =& - \sqrt{\frac{2} {3}}  A_1,
\label{eq:isorel03}
\end{eqnarray}
where $A_1$ ($A_0$) is the amplitude to produce the system $DK$
with an isospin quantum number equal to 1 (0). The $A_i$ amplitudes
in these formulae are reduced matrix elements,
in the terms of the Wigner-Eckart theorem, of the isoscalar
Hamiltonian.

A similar set of relations holds for charged B meson decays
\begin{eqnarray}
A(B^+ \rightarrow \Dz D^+ \Kz ) &=&  \frac{1} {\sqrt{6}}  A_1 -\frac{1} {\sqrt{2}}
A_0 \label{eq:isorelc1}\\
A(B^+ \rightarrow \Dz \Dzb K^+ ) &=& \frac{1} {\sqrt{6}} A_1 +\frac{1} {\sqrt{2}}
A_0 \label{eq:isorelc2}\\
A(B^+ \rightarrow D^-  D^+ K^+ )& =& - \sqrt{\frac{2} {3}}  A_1,
\label{eq:isorelc3}
\end{eqnarray}
where the $A$ amplitudes are the same as for the neutral B decays.
Identical equations hold for the other set of decays, \BDDsK, \BDsDK~and  \BDsDsK,
with different amplitudes $A$ in each case. Equivalent relations can be obtained
considering the isospin quantum numbers of different subsystems of the
final state ($D \Dbar$, $\Dbar K$). The $DK$ subsystem is
chosen here because in this case the transitions of Eqs.
(\ref{eq:isorel03}) and (\ref{eq:isorelc3}),  proceeding only
through the color-suppressed
diagrams of Fig.~\ref{Fi:diagrams} (left plot),  are
associated only to the $A_1$ amplitude.

\begin{figure}[htb]
\begin{minipage}{6cm}
\begin{flushleft}
\includegraphics[width=7cm]{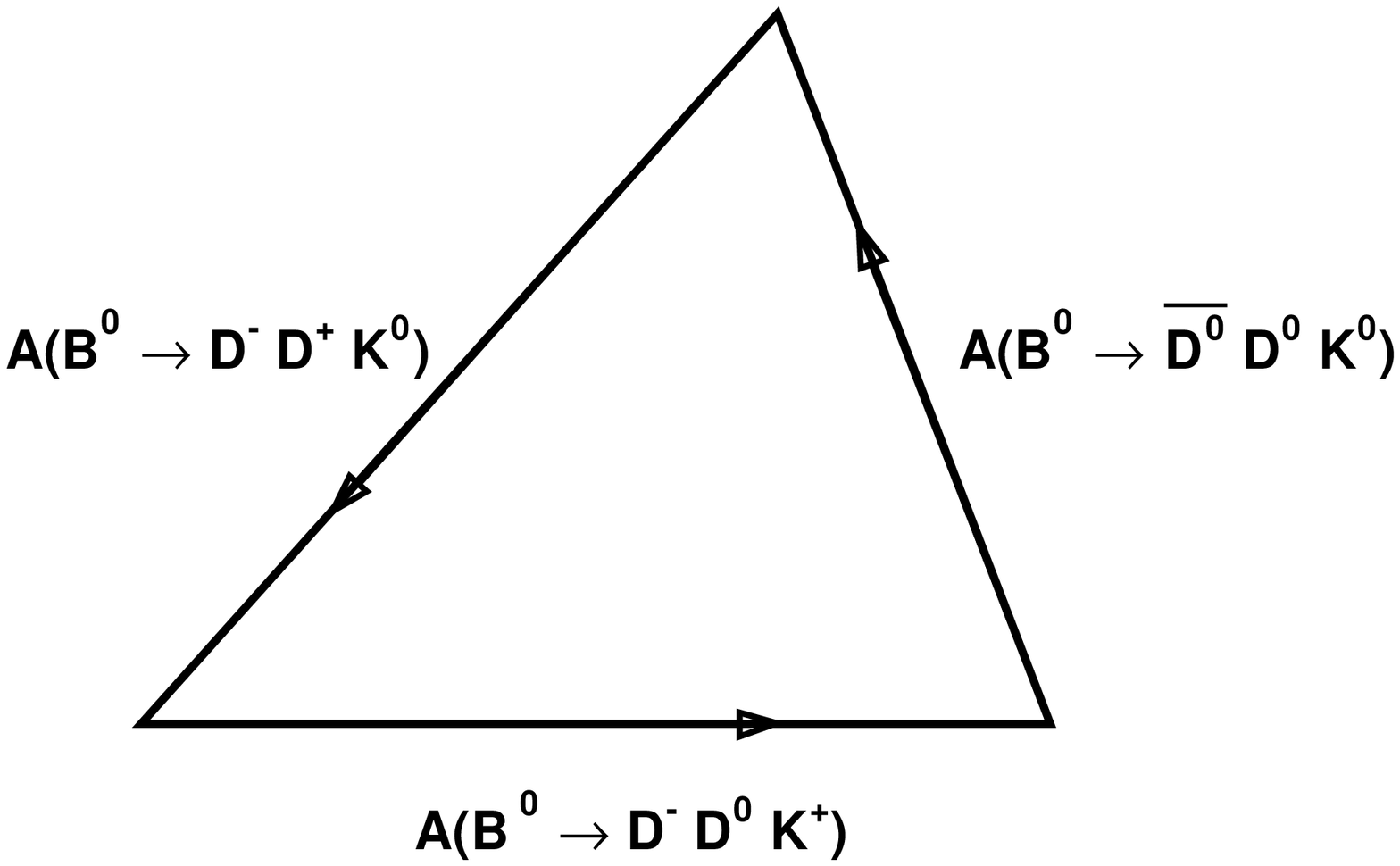}\\
\end{flushleft}
\end{minipage}
\hfill
\begin{minipage}{8cm}
\begin{flushright}
\includegraphics[width=7.cm]{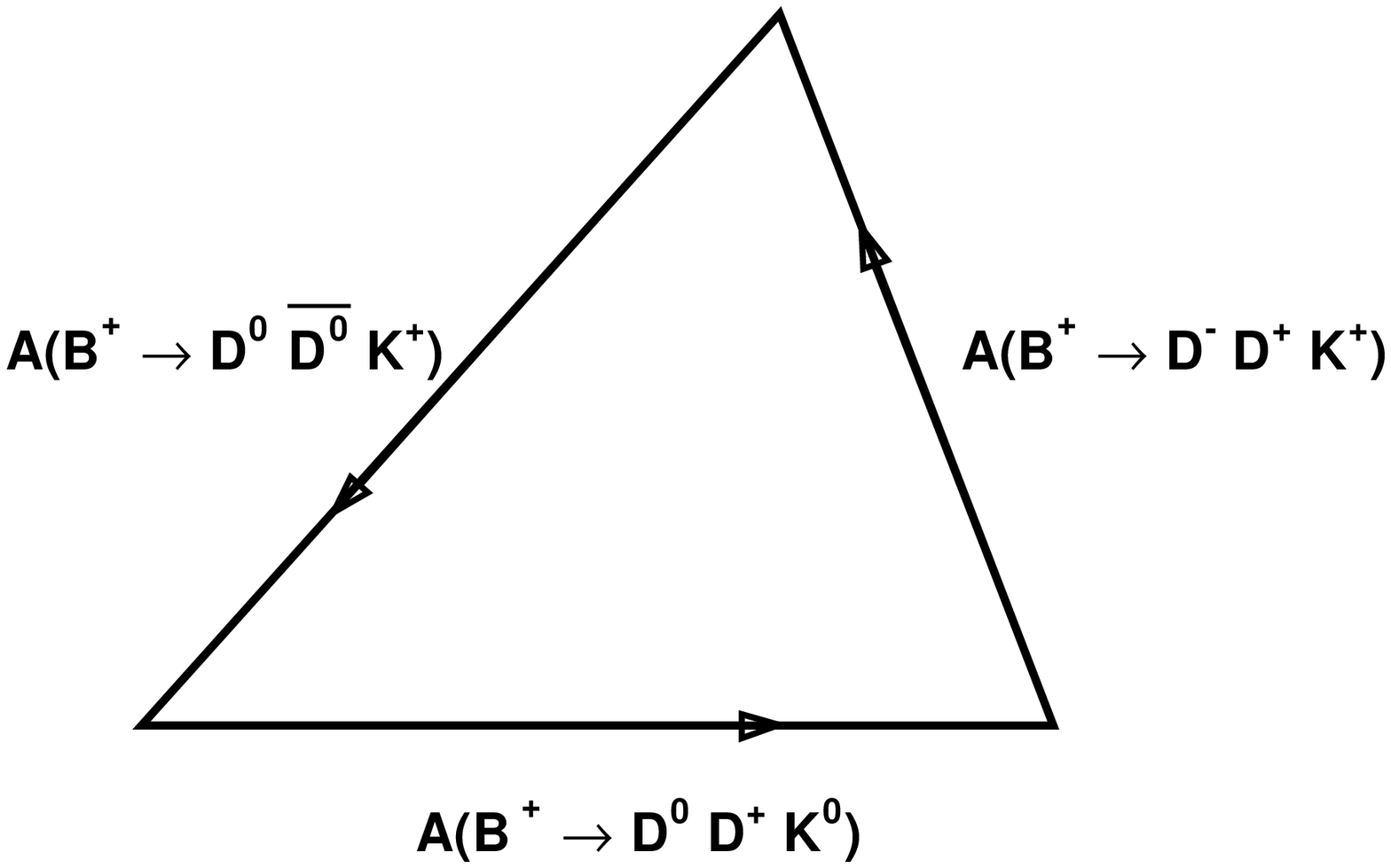}\\
\end{flushright}
\end{minipage}
\caption{Isospin triangles for the $\Bz$ (left) and $\Bp$ (right)
amplitudes.}
\label{fig:triangle_th}
\end{figure}

The relations presented above can be cast in the form of a triangle relation
between the amplitudes:
\begin{eqnarray}
-A(\Bz \rightarrow D^- \Dz K^+ ) &=& A(\Bz \rightarrow D^- D^+ \Kz ) + A(\Bz \rightarrow \Dzb  \Dz \Kz ) \label{eq:triangle_bz}\\
-A(B^+ \rightarrow \Dz D^+ \Kz ) &=& A(B^+ \rightarrow \Dz \Dzb K^+ ) +
A(B^+ \rightarrow D^-  D^+ K^+ ), \label{eq:triangle_bp}
\end{eqnarray}
which are depicted in Fig.~\ref{fig:triangle_th}.
The two triangles for $\Bz$ and $\Bp$ decays are identical according to the
isospin relations, however experimentally it is advantageous to build the
triangles separately with the $\Bz$ and $\Bp$ amplitudes.

We finally notice that Eqs.~(\ref{eq:isorel01}) to~(\ref{eq:triangle_bp})
are valid not only for the total decay amplitude but also for
each helicity amplitude separately as well as for
the amplitude as a function of the Dalitz plot coordinates. The amplitudes and phases we extract from the fit are averaged over the Dalitz plot as well as over all the accessible final states (vector polarizations, partial waves, ...).
This remark is of particular importance since it is now well known that many resonances are present in the $\BDDK$ decays, such as the $\Psi(3770)$, the $D_{s1}(2536)$, the $X(3872)$, and the $D_{s1}(2700)$ mesons~\cite{ref:reso}.

\section{Study of the experimental results}

The branching fractions for the charged and neutral $B$ meson decay can be written
\begin{eqnarray}
\BR(\Bp \rightarrow \Dbar^{(*)} D^{(*)} K) &=& \frac{\tau_+ }{(2 \pi)^3~32 M_{B^+}^3} \left ( \int dm^2_{\Dbar^{(*)} D^{(*)}}
dm^2_{D^{(*)} K} \right ) | A (\Bp \rightarrow \Dbar^{(*)} D^{(*)} K) |^2 ~~~~~~ \label{eq:branchingp}\\
\BR(\Bz \rightarrow \Dbar^{(*)} D^{(*)} K) &=& \frac{\tau_0}{(2 \pi)^3~32 M_{B^0}^3} \left (\int dm^2_{\Dbar^{(*)} D^{(*)}}
dm^2_{D^{(*)} K}\right ) | A (\Bp \rightarrow \Dbar^{(*)} D^{(*)} K) |^2 ,
\label{eq:branching0}
\end{eqnarray}
where $\tau_+ = 1.638 \times 10^{-12}~{\rm s}$ and
$\tau_0 = 1.525 \times 10^{-12}~{\rm s}$ \cite{ref:pdg}
are the lifetimes of the $\Bp$ and $\Bz$ mesons,
$M_{B^+}$ and $M_{B^0}$ are the masses of the $B^+$ and $B^0$
mesons, $m_{\Dbar^{(*)} D^{(*)}}$ and
$m_{D^{(*)} K}$ are the invariant masses of the ${\Dbar^{(*)} D^{(*)}}$ and
${D^{(*)} K}$ subsystems, and the integral is computed numerically
over the allowed region of the three-body phase space.

The BABAR Collaboration has recently studied the full set of $\BDDK$ decays
and has provided precise measurements for all these modes~\cite{ref:ddk2011}. We use also the experimental results from the Belle Collaboration~\cite{ref:belleDDKCP,ref:belleDDKDsJ} which are available for the modes \modex\ and \modexi. These two modes from Belle are combined with the corresponding ones from BABAR assuming fully correlated systematic uncertainties. Table~\ref{tab:ddkyields} presents the measurements of the \BDDK\ final states after having combined the BABAR and Belle results.

The BABAR and Belle data have been collected at the PEP-II and KEKB accelerators
from the reaction $e^+ e^- \rightarrow \Upsilon(4S) \rightarrow B \Bbar$.
To compute the branching fractions, it has been assumed that
$\BR(\Upsilon(4S) \rightarrow B^+ B^-) = \BR(\Upsilon(4S) \rightarrow \Bz
\Bzb) = 0.5 $. However these equalities do not necessarily hold.
 In order to account for
this factor, we rewrite Eqs.~(\ref{eq:branchingp}) and (\ref{eq:branching0})
in term of the rescaled amplitudes
\begin{equation}
\tilde{A}  = \frac{A}{\sqrt{2 \BR(\Upsilon(4S) \rightarrow \Bz \Bzb) }}.
\end{equation}
The expression for $\BR(\Bp \rightarrow \Dbar^{(*)} D^{(*)} K)$ is then multiplied
by the additional factor
\begin{equation}
f_{+/0}= \frac{\BR(\Upsilon(4S) \rightarrow B^+ B^-)}{\BR(\Upsilon(4S) \rightarrow \Bz \Bzb )}.
\end{equation}

The experimental data are fitted simultaneously using the $\chi^2$ method:
\begin{equation}
\chi^2 = (\BR_{\mathrm{exp}} - \BR_{\mathrm{pred}})^T V^{-1} (\BR_{\mathrm{exp}} - \BR_{\mathrm{pred}}) + \frac {(f_{+/0} -f^{\mathrm{WA}}_{+/0})^2} {\sigma_{f^{\mathrm{WA}}_{+/0}}^2},
\label{eq:chi2}
\end{equation}
where $\BR_{\mathrm{exp}}$ represents the vector of the branching fraction measurements, $\BR_{\mathrm{pred}}$ represents the vector of the branching fraction predictions, and the superscript $T$ denotes the transposed vector. The predictions depend on 13 parameters which are $f_{+/0}$ and, for each set of decays,
$  |\tilde{A_1}|$, $|\tilde{A_0}|$, and $ \delta = arg ( \tilde{A_1} \tilde{A_0^*} )$.
The matrix $V$ is the covariance matrix between the 22 branching fraction measurements, which allows to take properly into account the systematic uncertainties that are common and correlated between each mode. The correlated systematic uncertainties consist of
uncertainties originating from the signal shape, the reconstruction and the identification of particles (charged tracks, soft pions from $D^{*+}$ decays, \KS, $\pi^0$, single photon, and $K^+$ identification), the branching fractions of the secondary decays ($D^{(*)}$ and \KS), and the accounting of the number of $B\Bbar$ pairs produced in the experiment (see Table III of Ref.~\cite{ref:ddk2011}). We separate each contribution of these systematic effects in order to break down the problem into quantities which are completely independent or completely correlated. We sum these separate covariance matrices together to obtain the total covariance matrix, where the partial correlation structures emerge.
The last term in Eq.~(\ref{eq:chi2}) constrains
$f_{+/0} $ to the world average value $f^{\mathrm{WA}}_{+/0}=1.065 \pm 0.026$~\cite{ref:pdg}.

\begin{figure}[htb]
\begin{center}
\includegraphics[width=15cm]{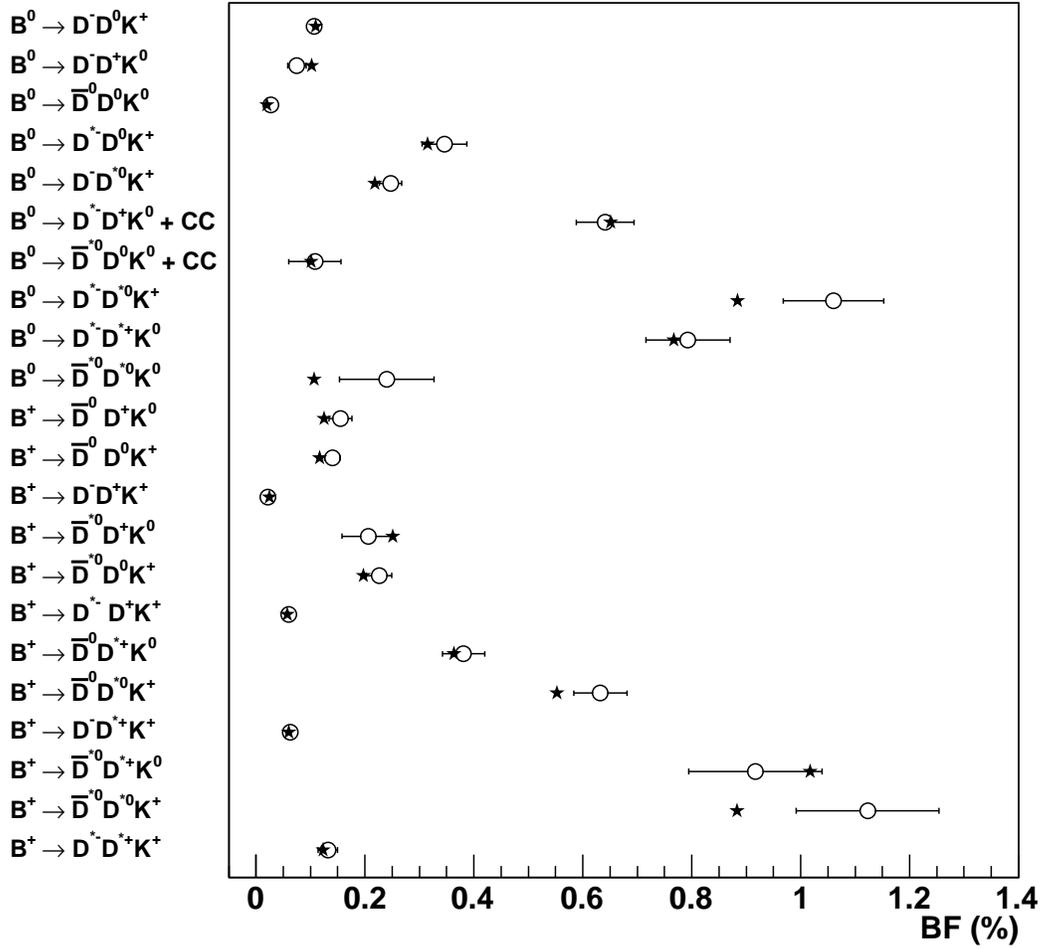}\\
\end{center}
\caption{Results of the $\chi^2$ fit to the experimental
branching fractions. The fitted branching fractions
are shown by the stars while the points with error bars show the
measured values.}
\label{fig:bfplot}
\end{figure}

\begin{figure}[htb]
\begin{center}
\includegraphics[width=15cm]{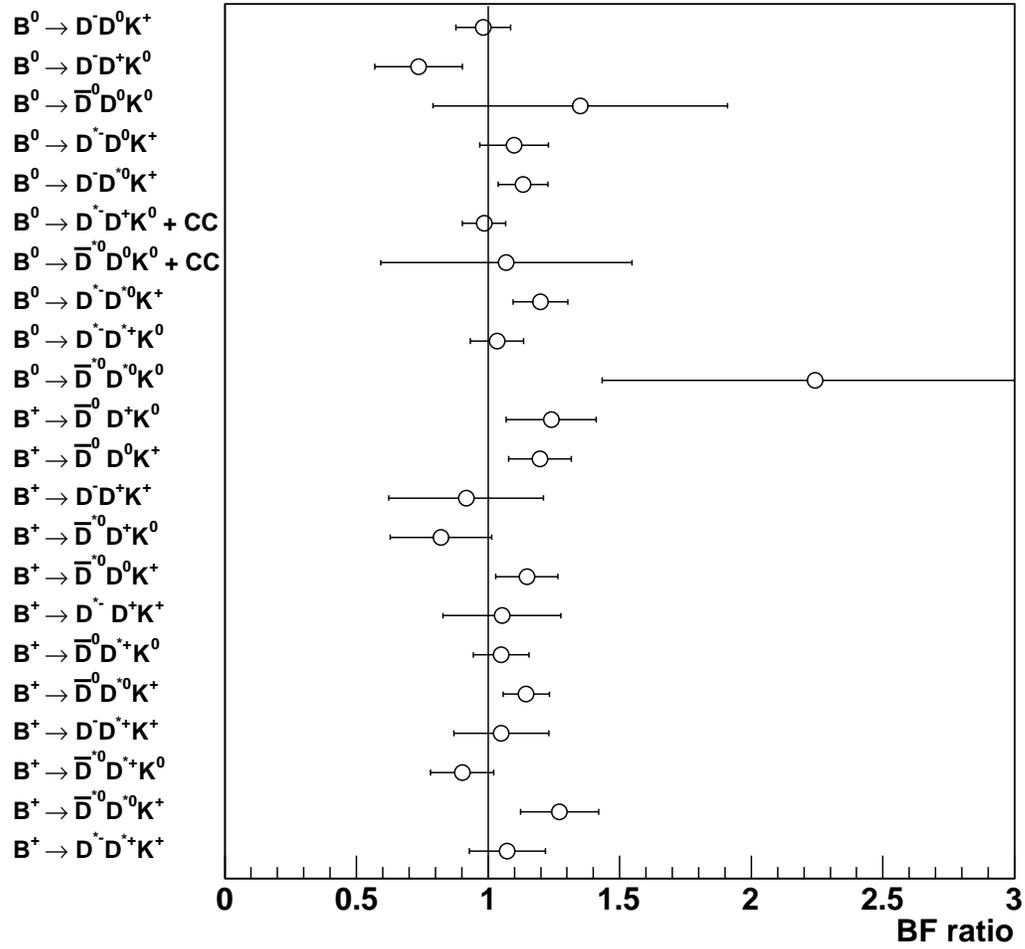}\\
\end{center}
\caption{Ratios of measured branching fractions over predicted branching fractions, $\BR_{\mathrm{exp}}/\BR_{\mathrm{pred}}$. The vertical line shows the case $\BR_{\mathrm{exp}}/\BR_{\mathrm{pred}}=1$.}
\label{fig:bfplot2}
\end{figure}

The results of the minimization of this $\chi^2$  are reported in Tables~\ref{tab:ddkyields}
and~\ref{table:fitres}. The overall
agreement between the measured and predicted branching fractions is fair
as can be judged from Table~\ref{tab:ddkyields}, Figs.~\ref{fig:bfplot} and \ref{fig:bfplot2}, and
from the value $\chi^2 = 18.9$ for 10 degrees of freedom ($n_{\mathrm{dof}}$) with a probability of 4.1\%.
We observe that the main source of the disagreement concerns the modes containing one or two $D^{*0}$ mesons, with a measured branching fraction systematically above the predicted value. This could point to a systematic shift that was not properly taken into account in the experimental analysis.
For some $\Bz$ decays which are not distinguishable experimentally, only the sum of the branching fraction with the charge conjugate final state has been measured. We present
in Table~\ref{table:BFunsummed} the fitted values for the individual
branching fractions.

The fit has also been conducted without the constraint on $f_{+/0}$.
We obtain a value
\begin{equation}
f_{+/0}=1.100 \pm 0.056
\end{equation}
which is in good agreement, while less precise, with the world average.

\begin{figure}[htb]
\begin{center}
        \includegraphics[width=15cm]{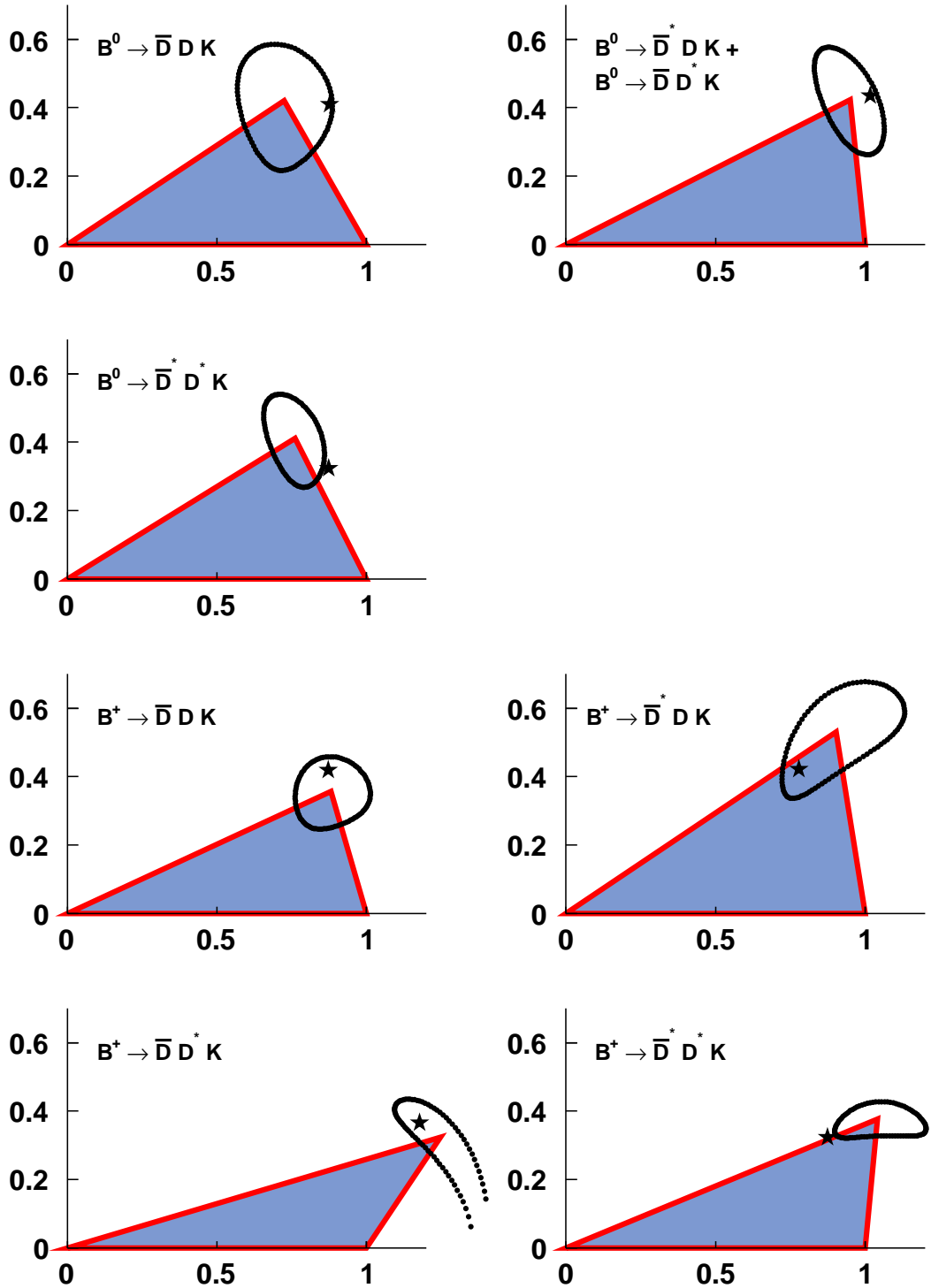}
\end{center}
\caption{Isospin triangles for the $\BDDK$ amplitudes.
Each plot presents the measured vertex of the triangle, where the
basis has been normalized to unity. The dotted contour shows the
one standard deviation region. The star shows the result of the fit.}
\label{fig:triangles_exp}
\end{figure}

An alternative way of displaying the experimental results and the fit results
is given by the isospin triangles introduced in the above.
For ease of comparison, we normalize the triangles to the size of
the basis
($|A(\Bz \rightarrow D^{(*)-} D^{(*)0} K^+ )|$ and
$|A(B^+ \rightarrow D^{(*)0}  D^{(*)+}  \Kz ) |$):
therefore the lower side extends in each case from (0,0) to (1,0) and the
shapes of the triangles can be directly compared. Given that we have only
a measurement of the sides, there is a fourfold ambiguity on the vertex of
the triangle. We choose consistently the same solution for its orientation.
The seven measured triangles defined in this way are shown in
Fig.~\ref{fig:triangles_exp} together with the fit result. We notice that
in all the cases the shape of the triangles presents large angles.

\begin{table*}[htb]
\caption{Branching fractions (\BR) for each $\BDDK$ mode. The second column shows the experimental results while the third column presents the result of the $\chi^2$  fit.
The first error on the experimental branching fraction is the statistical uncertainty
and the second is the systematic uncertainty~\cite{ref:ddk2011,ref:belleDDKCP,ref:belleDDKDsJ}. The experimental results from the modes \modex\ and \modexi\ are a combination between the BABAR and Belle measurements.}
\begin{center}
\begin{tabular}{|l|c|c|} \hline
\CellTop
  $B$ decay mode   & \BR\ exp. $(10^{-4})$ & \BR\ fit $(10^{-4})$   \\
\hline
\multicolumn{3}{|c|}{\CellTop \Bz decays through external \W-emission amplitudes}
\\ \hline
\CellTop
$\Bz\to \Dm \Dz\Kp$        & $10.7 \pm 0.7 \pm 0.9$ & 10.9 \\
$\Bz\to\Dm\Dstarz\Kp$       & $34.6 \pm 1.8 \pm 3.7$ &  31.5\\
$\Bz\to D^{*-} \Dz\Kp$      & $24.7 \pm 1.0 \pm 1.8 $ & 21.8\\
$\Bz\to D^{*-}\Dstarz\Kp$     & $106.0 \pm 3.3 \pm 8.6 $& 88.4 \\ \hline
\multicolumn{3}{|c|}{\CellTop \Bz decays through external+internal
\W-emission amplitudes} \\
\hline
\CellTop
$\Bz\to\Dm\Dp\Kz$             & $7.5 \pm 1.2 \pm 1.2 $ & $10.2
$\\
$\Bz\to D^{*-}\Dp\Kz+\Dm D^{*+}\Kz$   & $64.1 \pm 3.6 \pm 3.9 $ &  65.1
\\
$\Bz\to D^{*-} D^{*+}\Kz$     & $79.3\pm 3.8 \pm 6.7$ & 76.7 \\
\hline
\multicolumn{3}{|c|}{\CellTop \Bz decays through internal \W-emission
amplitudes}
 \\ \hline
\CellTop
$\Bz\to \Dzb \Dz \Kz$       & $2.7  \pm 1.0 \pm 0.5 $ & $2.0 $ \\
$\Bz\to\Dzb \Dstarz \Kz+ \Dstarzb \Dz \Kz$  & $10.8 \pm 3.2\pm 3.6$ & 10.1   \\
$\Bz\to \Dstarzb \Dstarz \Kz$  & $24 \pm 5.5 \pm 6.7 $ & $10.7 $ \\
\hline
\multicolumn{3}{|c|}{\CellTop \Bu decays through external \W-emission amplitudes}
 \\ \hline
\CellTop
$\Bu\to \Dzb \Dp\Kz$         & $15.5 \pm 1.7  \pm 1.3 $ & $12.5$ \\
$\Bu\to \Dzb D^{*+}\Kz$     &  $38.1 \pm 3.1 \pm 2.3 $ & $36.3$ \\
$\Bu\to \Dstarzb\Dp\Kz$    & $20.6 \pm 3.8 \pm 3.0 $ & 25.1  \\
$\Bu\to \Dstarzb D^{*+}\Kz$  & $91.7 \pm 8.3 \pm 9.0$ & 101.7\\
\hline
\multicolumn{3}{|c|}{\CellTop \Bu decays through external+internal
\W-emission amplitudes} \\
\hline
\CellTop
$\Bu\to \Dzb \Dz \Kp$       & $14.0 \pm 0.7 \pm 1.2 $       & 11.7 \\
$\Bu\to \Dzb \Dstarz\Kp$   & $63.2 \pm 1.9 \pm 4.5 $ & $55.2$ \\
$\Bu\to\Dstarzb\Dz\Kp$  & $22.6  \pm 1.6 \pm 1.7$       & 19.7 \\
$\Bu\to\Dstarzb\Dstarz\Kp$   & $112.3 \pm 3.6 \pm 12.6$ &88.3 \\
\hline
\multicolumn{3}{|c|}{\CellTop \Bu decays through internal \W-emission amplitudes}
 \\ \hline
\CellTop
$\Bu\to \Dm\Dp\Kp$ & $2.2 \pm 0.5 \pm 0.5$ & $2.4 $ \\
$\Bu\to\Dm D^{*+}\Kp$  & $6.3 \pm 0.9 \pm 0.6$ & $6.0$ \\
$\Bu\to D^{*-} \Dp\Kp$  & $6.0 \pm 1.0 \pm 0.8$ & 5.7      \\
$\Bu\to D^{*-} D^{*+}\Kp$  & $13.2 \pm 1.3 \pm 1.2$ & $12.3$ \\ \hline
\end{tabular}
\end{center}
\label{tab:ddkyields}
\end{table*}

\begin{table*}[tb]
\caption{Results of the $\chi^2$ fit to the experimental
branching fractions for the amplitudes and phases. The superscripts $LL$, $L*$, $*L$ and $**$
refer to the $\BDDKspec$, $\BDDsK$, $\BDsDK$ and  $\BDsDsK$ decays
respectively. The amplitude values are in  units of $10^{-5}$ while the
phases $\delta$ are in degrees.
}
\begin{center}
\begin{tabular}{|l|c|} \hline
Parameter & Value \\
\hline
\CellTop
$|A_1^{LL}|$ & $ 0.23 \pm 0.03 $ \\
$|A_0^{LL}|$ & $ 0.59 \pm 0.02 $ \\
$\delta^{LL}$ & $ 94 \pm 8 $  \\
\hline
\CellTop
$|A_1^{L*}|$ & $ 0.42 \pm 0.04 $ \\
$|A_0^{L*}|$ & $ 1.33 \pm 0.04 $\\
$\delta^{L*}$ & $ 53 \pm 9 $ \\
\hline
\CellTop
$|A_1^{*L}|$ & $ 0.41 \pm 0.04 $  \\
$|A_0^{*L}|$ & $ 0.92 \pm 0.03 $  \\
$\delta^{*L}$ & $ 103 \pm 7 $ \\
\hline
\CellTop
$|A_1^{**}|$ & $ 0.72 \pm 0.05 $ \\
$|A_0^{**}|$ & $ 2.28 \pm 0.08 $ \\
$\delta^{**}$ & $ 100 \pm 7 $ \\
\hline
$f_{+/0}$ & $ 1.071 \pm 0.023 $ \\
 \hline
 \CellTop
$\chi^2/ n_{\mathrm{dof}} $ & 18.9/10  \\
Prob$(\chi^2, n_{\mathrm{dof}}) $ & 4.1 \%  \\
 \hline
\end{tabular}
\end{center}
\label{table:fitres}
\end{table*}

\begin{table*}[htb]
\caption{Fitted values of the branching fractions
for the  $B^0 \to \Dbar D^* K^0$ and $B^0 \to \Dbar^* D K^0$
decays which have not been measured individually.}
\begin{center}
\begin{tabular}{|l|c|} \hline
\CellTop
$B$ decay mode & \BR\ fit $(10^{-4})$   \\
\hline
\CellTop
$\Bz\to D^{*-}\Dp\Kz $ & 17.1 \\
$\Bz\to \Dm D^{*+}\Kz$    &  48.0 \\
$\Bz\to \Dstarzb \Dz \Kz$ &  4.9 \\
$\Bz\to\Dzb \Dstarz \Kz$  &  5.2 \\
\hline
\end{tabular}
\end{center}
\label{table:BFunsummed}
\end{table*}

\clearpage

\section{Discussion}

\subsection{Dynamical features of the amplitudes}

The amplitudes and phases extracted from the data present
some distinctive features.
First, within each set, the amplitude related to the color-suppressed decays
is much smaller, as expected. The ratios $A_0/A_1$ are presented
in Table \ref{table:aratio}. These ratios are very close to the
na\"{\i}ve expectation
of a suppression factor $N_c=3$, where $N_c$ is the number of colors.

Second, the central values for the relative phases $\delta$ are
in all cases large and close to $90^\circ$.
From this we can conclude that there is a
firm indication for large strong phases in these amplitudes.
This suggests the presence of non-negligible Final State Interaction
for these decays. This is both an important indication {\it per se}
and has also consequences for the $CP$ violation studies that will be
discussed in the next section.

\begin{table*}[htb]
\caption{Ratios $A_0/A_1$ from the fit to the data. The uncertainties take into account the fit correlations between $A_0$ and $A_1$.}
\begin{center}
\begin{tabular}{|c|c|} \hline
ratio & value \\
\hline
$ |A_0^{LL}|/|A_1^{LL}|$ & $2.57 \pm 0.37 $  \\
$ |A_0^{L*}|/|A_1^{L*}|$ & $3.15 \pm 0.28 $ \\
$ |A_0^{*L}|/|A_1^{*L}|$ & $2.23 \pm 0.26 $ \\
$ |A_0^{**}|/|A_1^{**}|$ & $3.17 \pm 0.21 $ \\
\hline
\end{tabular}
\end{center}
\label{table:aratio}
\end{table*}

\subsection{Implications for the measurement of $\sin(2\beta)$ and $\cos(2\beta)$}

All the $\Bz \to \Dbar^{(*)} D^{(*)} \Kz$ final states are in principle good candidates
for the measurement of the $\beta$ angle of the unitarity matrix~\cite{ref:CPDDK_1,ref:CPDDK_2,ref:CPDDK_3}. The advantages of these modes, for example with respect to $\Bz \to \Dbar^{(*)} D^{(*)}$, are that they are Cabibbo-favored and present a small penguin contribution. Since both $B^0$ and $\Bzb$ can decay to $\Dbar^{(*)} D^{(*)} \Kz$, we expect a time-dependent $CP$ violating asymmetry. A study of the time-dependent Dalitz plot allows to access the phase $\beta$ related to the
$B^0$ and $\Bzb$ mixing. We notice that for $\Bz \to D^{*-} D^{*+} \Kz$, the
measured value of the branching fraction ($79.3 \pm 3.8 \pm 6.7 \times 10^{-4}$)
and the value predicted by our fit ($76.7 \times 10^{-4}$) are almost a factor
two lower that what was anticipated in Ref.~\cite{ref:CPDDK_3},
thereby  unfortunately also reducing the comparative advantage of this mode
with respect to $\Bz \to D^{*-} D^{*+}$.

The BABAR experiment did a study of the final state \modex\ in this context and was able to constrain $\cos 2\beta$ to be positive at the 94\% confidence level (under some theoretical and resonant substructure assumptions, and using $230 \times 10^6 B\Bbar$ pairs)~\cite{ref:babarDDKCP}. The Belle experiment did a similar analysis on the same final state with $449 \times 10^6 B\Bbar$ pairs and did a measurement of the $CP$ violation parameters, although the study did not allow to conclude on the sign of $\cos 2\beta$~\cite{ref:belleDDKCP}.

Unfortunately, up to now, no other $\Bz \to \Dbar^{(*)} D^{(*)} \Kz$ modes have been studied in the context of $CP$ violation. From the BABAR data ($429 \times 10^6 B\Bbar$)~\cite{ref:ddk2011}, we see that the final state $\Bz \to D^{*-} D^{+} \Kz + D^{-} D^{*+} \Kz$ is observed with a significance of $13\sigma$, where $\sigma$ is the standard deviation, which shows that a $CP$-violation analysis would be possible.

For $\Bz \to D^{-} D^{+} \Kz$, a value of $7.5 \pm 1.2 \pm 1.2 \times 10^{-4}$ is reported (with a $5\sigma$ significance). In this case too, the estimated value of Ref.~\cite{ref:CPDDK_2}
($90 \times 10^{-4}$) is a factor 12 above the measurement. However,
we stress that this channel is a good candidate for $CP$-violation studies
because of the nature of the final state with three pseudoscalar particles.
This will facilitate the angular analysis to determine the
helicity amplitudes.

Finally we notice that the $\Bz \to D^{*-} D^{+} \Kz$ and
$\Bz \to D^{-} D^{*+} \Kz$ decay modes lead to final states accessible
to both $\Bz$ and $\Bzb$. They can therefore be analyzed in the same
way as described in Ref.~\cite{ref:rhopi}. The strong phases play an
important role for this analysis as the time-dependent
$CP$-asymmetry amplitudes are proportional to $\sin(2 \beta \pm \delta')$,
where $\delta'$ is the strong phase difference between
$A(\Bz \to D^{-} D^{*+} \Kz)$ and $A(\Bzb \to D^{-} D^{*+} \Kz)$.
The possibly large values of the strong phases noticed in the above need to be taken
into account for any estimate of the sensitivities of this analysis.

\section{Conclusion}

We have presented an isospin analysis of the $\BDDK$ decays, based on recent and precise measurements of these final states. A fit was performed using the isospin relations between the different final states. We find a good agreement between the experimental values and the fitted values.
The isospin amplitudes exhibit several peculiar features like the presence of color-suppression and large relative phases.
We find a value of $\frac{\BR(\Upsilon(4S) \rightarrow B^+ B^-)} {\BR(\Upsilon(4S) \rightarrow \Bz \Bzb)}$ equal to $1.100 \pm 0.056$, in agreement with other determinations of this quantity. We have discussed the features of our result and showed the implications for $CP$-violation measurements using these decays.

\section{Acknowledgments}

The authors would like to warmly thank S\'ebastien Descotes-Genon for careful reading of this Letter and useful discussions.

\end{document}